\documentclass[floats,aps,showpacs,twocolumn,superscriptaddress]{revtex4-1} 

\usepackage[dvips]{graphicx}
\usepackage{amsfonts}
\usepackage{amsmath,amssymb}

\newcommand{\ud}{\text{d}}
\newcommand{\ui}{\text{i}}
\newcommand{\ue}{\text{e}}

\newcommand{\br}{{\bf r}}
\newcommand{\bx}{{\bf x}}

\newcommand{\bq}{{\bf q}}
\newcommand{\bp}{{\bf p}}

\newcommand{\qs}{q_{\perp}}
\newcommand{\qp}{q_{\parallel}}
\newcommand{\ps}{p_{\perp}}
\newcommand{\pp}{p_{\parallel}}
\newcommand{\rs}{r_{\perp}}
\newcommand{\rp}{r_{\parallel}}

\newcommand{\Imag}{\text{Im}\,}
\newcommand{\Le}{L}
\newcommand{\Ar}{A}

\newcommand{\Lreg}{L_{\text{reg}}}
\newcommand{\Lch}{L_{\text{ch}}}
\newcommand{\Areg}{A_{\text{reg}}}
\newcommand{\Ach}{A_{\text{ch}}}
\newcommand{\wreg}{w_{\text{reg}}}
\newcommand{\wch}{w_{\text{ch}}}

\newcommand{\Nrot}{\bar{N}_{\text{rot}}}
\newcommand{\Nosc}{\bar{N}_{\text{osc}}}
\newcommand{\Lrot}{L_{\text{rot}}}
\newcommand{\Losc}{L_{\text{osc}}}
\newcommand{\Arot}{A_{\text{rot}}}
\newcommand{\Aosc}{A_{\text{osc}}}
\newcommand{\db}{\bar{d}}
\newcommand{\Nb}{\bar{N}}

\begin{document}

\title{Partial Weyl Law for Billiards}

\author{Arnd B\"acker}
\affiliation{Institut f\"ur Theoretische Physik, Technische Universit\"at
             Dresden, 01062 Dresden, Germany, EU}
\affiliation{Max-Planck-Institut f\"ur Physik komplexer Systeme, 
             N\"othnitzer Stra\ss{}e 38, 01187 Dresden, Germany, EU}

\author{Roland Ketzmerick}
\affiliation{Institut f\"ur Theoretische Physik, Technische Universit\"at
             Dresden, 01062 Dresden, Germany, EU}
\affiliation{Max-Planck-Institut f\"ur Physik komplexer Systeme, 
             N\"othnitzer Stra\ss{}e 38, 01187 Dresden, Germany, EU}

\author{Steffen L\"ock}
\affiliation{Institut f\"ur Theoretische Physik, Technische Universit\"at
             Dresden, 01062 Dresden, Germany, EU}

\author{Holger Schanz}
\affiliation{Institut f\"ur Maschinenbau, Hochschule Magdeburg-Stendal, 
             39114 Magdeburg, Germany, EU}
\affiliation{Max-Planck-Institut f\"ur Physik komplexer Systeme, 
             N\"othnitzer Stra\ss{}e 38, 01187 Dresden, Germany, EU}

\date{\today}

\begin{abstract}
For two-dimensional quantum billiards we derive the partial Weyl law, i.e.\ 
the average density of states, for a subset of eigenstates concentrating on an
invariant region $\Gamma$ of phase space. 
The leading term is proportional to the area of the billiard times 
the phase-space fraction of $\Gamma$.
The boundary term is proportional to the fraction of the boundary 
where \emph{parallel} trajectories belong to $\Gamma$.
Our result is numerically confirmed for the mushroom 
billiard and the generic cosine billiard,
where we count the number of chaotic and regular 
states, and for the elliptical billiard, 
where we consider rotating and oscillating states.
\end{abstract}

\maketitle
\noindent

Waves restricted to finite resonators in two or three dimensions have found
abundant applications in contemporary physics. Examples include
electromagnetic and acoustic resonators, microdisk lasers, atomic matter
waves in optical billiards, and quantum dots 
\cite{Sto1999,NoeSto1997,FriKapCarDav2001,ArcBaeCuaPriMenHer1998,
MarRimWesHopGos1992}. 
The average density of states is an essential observable 
of a resonator and dictates many physical
properties. Investigations of this quantity go back to Lord Rayleigh more than a
hundred years ago \cite{Ray1877} and have been a continuing topic of interest
ever since \cite{Wey1911,Wey1913,Kac1966,BalBloComb,BalHil1976,BerHow1994,
SiePriSmiUssSch1995,AreNitPetSte2009}. 
Today the fundamental result for the average density of states
\begin{equation}
\label{eq:dos_2d_billiard}
 \db(E) = \frac{A}{4\pi} - \frac{L}{8\pi} \frac{1}{\sqrt{E}} + \dots
\end{equation}
is known as Weyl's law, as he gave the first proof of the leading term 
\cite{Wey1911}. Equation~\eqref{eq:dos_2d_billiard} is formulated 
for the case of a two-dimensional quantum billiard with area $A$,
boundary length $L$, and Dirichlet boundary
conditions (in units $\hbar=2m=1$). Analogous results
for three dimensions, other types of waves and boundary conditions are 
available \cite{AreNitPetSte2009}, and for open systems a fractal Weyl law 
was proposed \cite{LuSriZwo2003}.

The first term of eq.~\eqref{eq:dos_2d_billiard} depends on the area $A$ of
the billiard only. As for any quantum system
it is obtained by counting the number of Planck cells in 
the phase space available at energy $E$. The
second term was already conjectured by Weyl \cite{Wey1913}.
It is specific for billiards or
resonators and depends on the length $L$ of the boundary. As the wave function
must vanish on the boundary, a layer is depleted which has a length $L$ and a 
width of the order of the wave length $\lambda\sim 1/\sqrt{E}$.
Semiclassically, the second 
term can be interpreted as a contribution from closed
trajectories which are reflected
perpendicularly at the boundary 
\cite{BalBloComb,SieSmiCreLit1993,Hes1994}.
Higher order corrections in eq.~\eqref{eq:dos_2d_billiard} arise, e.g.,
due to corners and curvature effects \cite{Kac1966,BalBloComb,
BalHil1976,BerHow1994,SiePriSmiUssSch1995,AreNitPetSte2009}.

\begin{figure}[tb]
  \begin{center}
     \includegraphics[width=85mm]{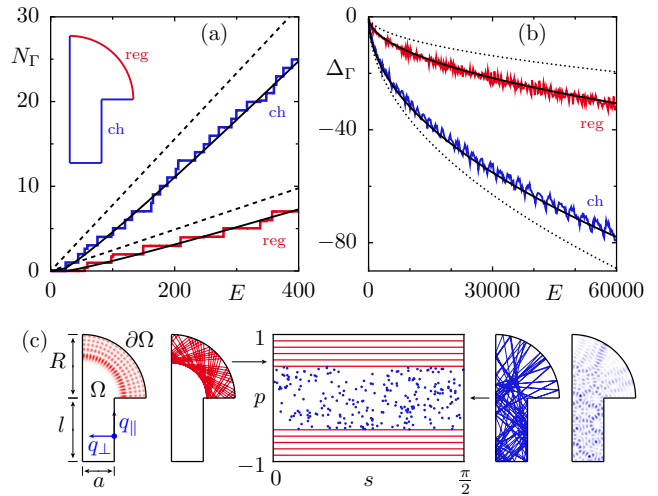}
 \caption{(Color online) (a) Spectral staircase $N_\Gamma(E)$
           for regular and chaotic eigenstates of the 
           desymmetrized mushroom billiard with $R=1$, $l=1$, and $a=0.5$. 
           We compare numerical data
           with the first term (dashed lines) and both terms 
           of eq.~\eqref{eq:N_2d_billiard_gamma} (smooth solid lines).
           The inset shows the regular and chaotic parts of the boundary.
           (b) Same data after subtracting the area term, 
           $\Delta_\Gamma(E) = N_\Gamma(E)-A_\Gamma E/(4\pi)$, 
           compared to the second term of eq.~\eqref{eq:N_2d_billiard_gamma} 
           (smooth solid lines) shown over a larger energy range.               
           Determining $L_\Gamma$ in 
           eq.~\eqref{eq:N_2d_billiard_gamma} from perpendicular instead of 
           parallel trajectories gives an incorrect result (dotted lines).
           (c) Phase space at the circular boundary 
           with regular (lines) and chaotic (dots) regions and 
           illustrations of trajectories and eigenfunctions.}
 \label{fig:intro}
\end{center}
\end{figure}

Generic billiards, which are studied for electromagnetic, acoustic, and 
matter waves \cite{Sto1999,NoeSto1997,FriKapCarDav2001,
ArcBaeCuaPriMenHer1998}, have a phase space containing 
several dynamically separated
domains, such as regular and chaotic regions, see fig.~\ref{fig:intro}(c). 
The spectrum consists of sub-spectra with eigenfunctions mainly 
concentrating on one of these invariant regions $\Gamma_i$, according to the 
semiclassical eigenfunction hypothesis \cite{Per1973,Ber1977,Vor1979}. 
It is a fundamental question to know the corresponding partial average 
density of states $\db_{\Gamma_{i}}(E)$, 
where $\db(E)=\sum_{i}\db_{\Gamma_{i}}(E)$. 
This is essential for studying the spectral statistics 
\cite{Haa2000} of such sub-spectra. The partial density of 
states is also compulsory for the determination of transition rates with 
Fermi's golden rule \cite{BaeKetLoeRobVidHoeKuhSto2008}.
Furthermore, it is required when an external 
coupling to the system is not uniform in
phase space, e.g., the total internal reflection in optical resonators
\cite{HacVivElaHaa2001,Sto2001} or tilted leads attached to a quantum dot
\cite{BirAkiFerVasCooAoySug1999}.

In this paper we derive the partial Weyl law,
\begin{equation}
 \label{eq:dos_2d_billiard_gamma}
  \db_\Gamma(E) = \frac{A_\Gamma}{4\pi} - \frac{L_\Gamma}{8\pi}
                 \frac{1}{\sqrt{E}},
\end{equation}
for a subset of eigenstates corresponding to an invariant region $\Gamma$ 
of phase space,
using the Wigner-Weyl transformation of the Green function.
As expected, for the area $A_\Gamma$ the ratio $A_\Gamma/A$ is the fraction 
of phase space occupied by $\Gamma$, see eq.~\eqref{eq:result_A_1}.
For the length $L_\Gamma$ we find that the ratio 
$L_\Gamma/L$ is the fraction of the billiard boundary where \emph{parallel} 
trajectories belong to $\Gamma$, see eq.~\eqref{eq:result_L_1}. 
This is unexpected as semiclassically the boundary term in 
eq.~\eqref{eq:dos_2d_billiard} originates from 
trajectories perpendicular to the boundary
\cite{BalBloComb,SieSmiCreLit1993,Hes1994}.
We confirm the result eq.~\eqref{eq:dos_2d_billiard_gamma} numerically 
for the mushroom billiard and the generic cosine billiard,
where we predict the number of regular and chaotic states, and for the
elliptical billiard, where we consider rotating and oscillating states.

Before we derive eq.~\eqref{eq:dos_2d_billiard_gamma} we exemplify its 
application for the desymmetrized mushroom billiard \cite{Bun2001} shown in
fig.~\ref{fig:intro}(c).  It is characterized by the radius $R$ of
the quarter circular cap, the stem width $a$, and the stem height $l$.
Classically one has two distinct phase-space regions visualized in
fig.~\ref{fig:intro}(c). All trajectories which are located only in the cap
of the mushroom are regular, while those entering the stem are chaotic
\cite{Bun2001}.
For the chaotic region eq.~\eqref{eq:result_A_1} yields $\Ach = \pi R^2/4
-[R^2\arcsin(a/R)+a\sqrt{R^2-a^2}]/2$ \cite{BaeKetLoeRobVidHoeKuhSto2008}.
For the length $\Lch$ we have to consider those parts of the billiard boundary
$\partial\Omega$ for which parallel trajectories belong to the chaotic region
(including the marginally stable bouncing-ball orbits). This
is the case for the straight boundaries, such that $\Lch =
2R+2l$ is the length of the chaotic boundary of the mushroom billiard. 
$\Areg$ and $\Lreg$ follow from $\Areg=\Ar-\Ach$ and
$\Lreg=\Le-\Lch=\pi R/2$, respectively. (For the full mushroom 
billiard one finds $\Lreg=\pi R$ and $\Lch=2R+2l$.)
In order to verify this prediction of the partial density of states, 
eq.~\eqref{eq:dos_2d_billiard_gamma}, we numerically solve
the time-independent Schr\"odinger equation, 
$-\Delta \psi_l(\bq) = E_l\psi_l(\bq)$,
for the desymmetrized mushroom billiard with Dirichlet boundary condition 
($\psi_l(\bq)=0$, $\bq\in\partial\Omega$), $R=1$, $l=1$, and $a=0.5$. 
We calculate the first $6024$ eigenstates $\psi_l$ 
using the improved method of particular solutions \cite{BetTre2005}. 
They can be classified as mainly regular or mainly
chaotic, depending on the phase-space region on which they concentrate
(fig.~\ref{fig:intro}(c)). There are several methods for this 
classification which give similar results. 
Here we determine the regular fraction $\wreg^l$ 
of an eigenstate $\psi_l$ of the mushroom by its projection on a basis of the 
regular region. For this basis we use the eigenstates 
$\psi_\text{qc}^{mn}$ of a quarter circle of radius $R=1$ with energy 
$E_{mn}$ and angular momentum $m>a\sqrt{E_{mn}}$.
These basis states are given by
$\psi_\text{qc}^{mn}(r,\varphi)=N_{mn}J_{m}(j_{mn}r)\sin(m\varphi)$, where
$m=2,4,\dots$ is the angular quantum number, $n=1,2,\dots$ is the radial
quantum number, $J_m$ is the $m$th Bessel function of the first kind,
$j_{mn}$ is the $n$th root of $J_m$, $E_{mn}=j_{mn}^{2}$, and
$N_{mn}=\sqrt{8/\pi}/J_{m-1}(j_{mn})$ is a normalization constant.
The projection of $\psi_l$ onto these basis states leads to the regular fraction 
$\wreg^l = \sum_{m,n} |\langle \psi_\text{qc}^{mn}|\psi_l\rangle|^2\,
\Theta(m-a\sqrt{E_{mn}})$ with $0 \leq \wreg^l \leq 1$. The chaotic fraction 
is then given by $\wch^l = 1-\wreg^l$. 
From $\wreg^l$ and $\wch^l$ the densities of the regular and the
chaotic states can be computed,
$d_{\Gamma}(E)=\sum_{l}w_{\Gamma}^l\,\delta(E-E_{l})$. 
For comparison with numerics we use the more convenient spectral staircase
function $N_{\Gamma}(E)=\int_{0}^{E}\ud\eta\,d_{\Gamma}(\eta)$. 
It has a step of size $w_{\Gamma}^l$ at eigenenergy $E_l$.  From
eq.~\eqref{eq:dos_2d_billiard_gamma} one finds
\begin{equation}
\label{eq:N_2d_billiard_gamma}
 \Nb_\Gamma(E) = \frac{A_\Gamma}{4 \pi}E - \frac{L_\Gamma}{4 \pi}\sqrt{E}.
\end{equation}
Figure~\ref{fig:intro}(a) shows the regular and the chaotic spectral staircase
for the mushroom billiard. We find excellent agreement with our prediction,
eq.~\eqref{eq:N_2d_billiard_gamma} (smooth solid lines). In
fig.~\ref{fig:intro}(b) we demonstrate that the boundary contribution of
eq.~\eqref{eq:N_2d_billiard_gamma} is in agreement with the
difference of the numerical data and the first term of
eq.~\eqref{eq:N_2d_billiard_gamma}. Using the semiclassical 
interpretation of the boundary term in eq.~\eqref{eq:dos_2d_billiard} 
one would naively expect that $L_\Gamma/L$ is the fraction of the 
boundary where perpendicular trajectories belong to $\Gamma$. However, 
this does not reproduce the data (dotted lines).
The numerical fluctuations arise due to oscillatory contributions to the density 
of states which are not considered here.
We have confirmed that under variation of the width $a$ of the stem 
of the mushroom the prediction eq.~\eqref{eq:N_2d_billiard_gamma} 
agrees with numerics.

Now we turn to the derivation of eq.~\eqref{eq:dos_2d_billiard_gamma}. 
The most general method to represent quantum states in phase space is 
their Wigner distribution. Other options, such as the Husimi distribution or
numerical methods for specific geometries as in the example above,
yield similar results and can be obtained from averages over the
Wigner distribution. In the semiclassical limit, according to the
semiclassical eigenfunction hypothesis, the weights $w^{l}_{\Gamma}$
for an invariant region $\Gamma$ of non-zero measure are either zero or
one and independent of the projection method.

Based on the Wigner distribution of an eigenstate $\psi_l$,
\begin{equation}
\label{eq:wigner}
 W_{l}(\bq,\bp)  = \frac{1}{\pi^2}\int \ud^2\br \,\ue^{2\ui\bp\br}
                   \psi_{l}^{*}(\bq+\br)\psi_{l}(\bq-\br),
\end{equation}
we define the phase-space resolved density of states for a phase-space point
$(\bq,\bp)$
\begin{equation}
\label{eq:density_states_rk}
  d_{\bq,\bp}(E)  =  \sum_{l=1}^\infty \delta(E-E_l) \, W_{l}(\bq,\bp).
\end{equation}
For an arbitrary region $\Gamma$ of phase space the density of states is 
given by the integral 
\begin{equation}
\label{eq:density_states_rk_int}
 d_{\Gamma}(E)=\int_{\Gamma} \ud^{2}\bq\, \ud^{2}\bp\, d_{\bq,\bp}(E).
\end{equation}
Using 
\begin{equation}
 \Imag G(\bx,\bx',E) = -\pi \sum_{l=1}^{\infty}\psi_{l}^{*}(\bx')\psi_l(\bx)\delta(E-E_l)
\end{equation}
eq.~\eqref{eq:density_states_rk} can be rewritten as
\begin{equation}
\label{eq:density_states_rk_gf}
d_{\bq,\bp}(E) = - \frac{1}{\pi^3} \int \ud^{2}\br \, 
                   \ue^{2\ui\bp\br} \,\Imag G(\bq-\br,\bq+\br,E),
\end{equation}
where $G(\bx,\bx',E)$ is the energy dependent Green function.
For billiards it satisfies
\begin{equation}
\label{eq:sgl_gf}
(\Delta+\kappa^{2})G(\bx,\bx',E)=\delta(\bx-\bx'),\quad \bx,\bx'\in\Omega,
\end{equation}
where $\kappa=\sqrt{E}$. On
$\partial\Omega$ it satisfies the boundary condition of the billiard
and is zero outside $\Omega$. 
Close to the boundary the curvature can be neglected and the Green function 
of a half-plane is an appropriate approximation. It is then given 
in the upper half-plane by
\begin{equation}
\label{eq:gf_2}
G(\bx,\bx',E) \approx  
\frac{1}{4\ui}[H_{0}^{+}(\kappa|\bx-\bx'|)-
  H_{0}^{+}(\kappa|\bx-\hat\bx'|)].
\end{equation}
Here $H_{0}^{+}$ denotes the Hankel function of the first kind and
$\hat\bx'$ is the mirror image of $\bx'$. 

Instead of requiring that the Green function vanishes in the lower half-plane, 
it is more appropriate to continue the Green function antisymmetrically across 
the boundary, i.e.\ eq.~\eqref{eq:gf_2} is extended to the full plane. In this 
way the corresponding Wigner function is adapted to the Dirichlet boundary
condition and yields a more faithful momentum distribution
\cite{FOOTNOTE}. 
Far from the boundary this modification has no effect and reduces to the 
standard definition.

Using polar coordinates for the momentum
$\bp=(p,\beta)$ and Cartesian coordinates for the position $\bq=(\qp,\qs)$, measured 
parallel and perpendicular to the boundary $\partial\Omega$, we
evaluate in eq.~\eqref{eq:density_states_rk_gf} the integrals over 
$\br$ and in eq.~\eqref{eq:density_states_rk_int} the integral over $p$, 
\begin{eqnarray}
\label{eq:d_dbqbeta}
 \db_{\bq,\beta}(E) & = & \frac{1}{4\pi^3} \int_{0}^{\infty} \ud p\,p
                      \int \ud^2 \br\, \ue^{2\ui\bp\br}\nonumber\\
                    &   & \times \left[J_0(2\kappa|r|)-J_0\left(2\kappa\sqrt{\rp^2+\qs^2}\right)\right],
\end{eqnarray}
where $J_0$ is the Bessel function of the first kind. For the $\br$-integration 
of the first Bessel function we use polar coordinates $\br = (r,\varphi)$, 
where $\varphi$ is the angle with respect to $\bp$, 
$\oint\ud\varphi\, \ue^{2\ui pr\cos\varphi} = 2\pi J_0(2pr)$, and 
$\int_{0}^{\infty} \ud r\, r\, J_0(2pr)J_0(2\kappa r)=\delta(p-\kappa)/4p$.
For the $\br$-integration of the second Bessel function in 
eq.~\eqref{eq:d_dbqbeta} we use Cartesian coordinates $\br=(\rp,\rs)$ leading
to $\bp\br = p\rp\cos(\beta-\beta_{\parallel}) + p\rs\sin(\beta-\beta_{\parallel})$,
where $\beta_{\parallel}(\bq)$ is the angle of the tangent 
to the boundary at $\bq\in\partial\Omega$.
Integration over $\rs$ gives $\pi\delta(\sin[\beta-\beta_{\parallel}])/p$ and 
integration over $p$ leads to $\pi\delta(\rp)/2$. Finally we obtain
\begin{equation}
\label{eq:res_d}
\db_{\bq,\beta}(E) = \frac{1}{8\pi^{2}} - \frac{1}{8\pi}
                   \delta(\sin[\beta-\beta_{\parallel}])J_0(2\kappa\qs).
\end{equation}
The $\delta$-function selects trajectories with momentum parallel to the 
boundary. According to eq.~\eqref{eq:density_states_rk_int} we have
\begin{equation}
\label{eq:int_res_d}
 \db_\Gamma(E) = \int_\Omega \ud^2 \bq \int_{0}^{2\pi} \ud\beta\, 
                  \chi_\Gamma(\bq,\beta)\,\db_{\bq,\beta}(E),
\end{equation}
where $\chi_\Gamma(\bq,\beta)$ is the characteristic
function of the phase-space region $\Gamma$. It is one when the trajectory
running at angle $\beta$ through the point $\bq$ belongs to $\Gamma$, 
zero if this is not the case, and $1/2$ on the boundary of $\Gamma$.
We now evaluate eq.~\eqref{eq:int_res_d} in the semiclassical limit 
$\kappa\to\infty$, where $J_0(2\kappa\qs) \to \delta(\qs)/\kappa$.
This gives the final result eq.~\eqref{eq:dos_2d_billiard_gamma} with
\begin{eqnarray}
\label{eq:result_A_1}
  A_\Gamma & = & \int_\Omega \ud^2\bq \, \frac{1}{2\pi} 
                 \int_0^{2\pi} \ud\beta \, \chi_\Gamma(\bq,\beta),\\
\label{eq:result_L_1}
  L_\Gamma & = & \oint_{\partial\Omega} \ud s\, 
                 \chi_\Gamma(\bq(s),\beta_{\parallel}(s)).
\end{eqnarray}
Here, $s$ is the arc length along the boundary, $\bq(s)$ is the corresponding
point on the boundary, and $\beta_{\parallel}(s)=\beta_{\parallel}(\bq(s))$ 
is the angle of the tangent to the boundary at that point. 
We thus find that $L_\Gamma/L$ is the fraction of the billiard boundary 
where parallel trajectories belong to $\Gamma$. 
Strictly speaking, $\chi_\Gamma(\bq(s),\beta_{\parallel}(s))$ in 
eq.~\eqref{eq:result_L_1} is obtained as a limit from trajectories starting at 
$\qp=s$ with $\beta\to\beta_\parallel$ and $\qs\to 0$.
This infinitesimal neighborhood has to be considered when the parallel 
trajectory with $\beta=\beta_{\parallel}$ and $\qs=0$ cannot be assigned to one 
of the regions $\Gamma$.
Note, that $A_\Gamma$ is 
proportional to the volume of $\Gamma$, while there is no such simple relation 
for $L_\Gamma$. For the special case where $\Gamma$ is the entire 
phase space we have $\chi_\Gamma\equiv 1$ leading to $A_\Gamma=A$ and 
$L_\Gamma=L$, such that eq.~\eqref{eq:dos_2d_billiard_gamma} reduces to 
eq.~\eqref{eq:dos_2d_billiard}.

We now give two explanations for the boundary term in 
eq.~\eqref{eq:dos_2d_billiard_gamma} and its relation to trajectories of 
$\Gamma$ that are parallel to the boundary:
(i) If $\Gamma$ is the entire phase space 
eqs.~\eqref{eq:density_states_rk_int} and \eqref{eq:density_states_rk_gf} 
lead to $d_{\bq}(E) = -\Imag G(\bq,\bq,E)/\pi$. The semiclassical contributions
to this Green function are given by closed trajectories which start and
end at $\bq$ \cite{Sto1999}. 
According to the second term in eq.~\eqref{eq:gf_2} they have
a length $2\qs$ corresponding to a perpendicular reflection at the boundary. 
This leads to the term $J_0(2\kappa \qs)/(4\pi)$, which
agrees with the second term in eq.~\eqref{eq:res_d} when integrated over
all $\beta$. If $\Gamma$ is a region in phase space, one needs the phase-space 
resolved density of states $d_{\bq,\bp}(E)$. 
Equation~\eqref{eq:density_states_rk_gf} shows that the direction of the 
momentum $\bp$ is not related to the direction of the trajectories which
semiclassically contribute to the Green function. Evaluating 
eq.~\eqref{eq:d_dbqbeta} leads to $\rp=0$, such that still 
perpendicularly reflected trajectories of length $2\qs$ give the boundary term. 
However, this contribution arises only if the direction $\beta$ 
of the momentum $\bp$ is parallel to the boundary. For the other directions 
the contribution cancels due to the phase factor $\exp(2\ui\bp\br)$ in 
eq.~\eqref{eq:d_dbqbeta}. 
(ii) An intuitive explanation of the boundary term can be given in terms of a 
plane wave, $\exp(\ui\pp\qp)\sin(\ps\qs)$, with Dirichlet boundary condition 
at $\qs=0$. Here the sine-term suppresses waves with $\ps=0$ and thus reduces
the density of states for these waves, which have a momentum $\bp$ parallel to 
the boundary. Semiclassically such waves correspond to trajectories
parallel to the boundary, in agreement with our result for the boundary term,
eq.~\eqref{eq:result_L_1}.

\begin{figure}[b]
  \begin{center}
     \includegraphics[width=85mm]{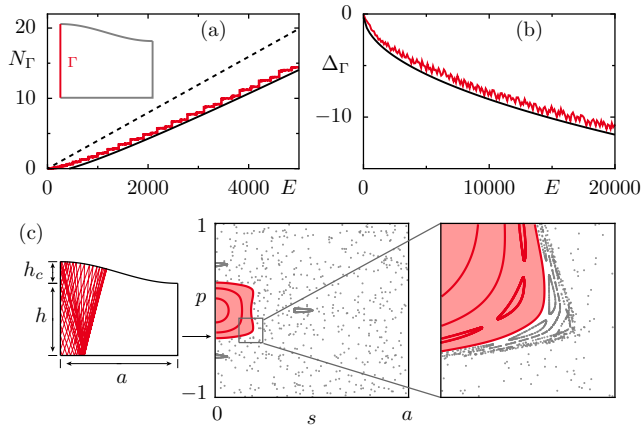}
 \caption{(Color online) (a) Spectral staircase $N_\Gamma(E)$
          for eigenstates concentrated in an invariant region $\Gamma$ of the 
          desymmetrized cosine billiard with $a=1.3$, $h=0.8$, and $h_c=0.24$. 
          We compare numerical data
          with the first term (dashed line) and both terms 
          of eq.~\eqref{eq:N_2d_billiard_gamma} (smooth solid line).
          The inset shows the part of the boundary corresponding to $\Gamma$.
          (b) Same data after subtracting the area term, 
          $\Delta_\Gamma(E) = N_\Gamma(E)-A_\Gamma E/(4\pi)$, 
          compared to the second term of eq.~\eqref{eq:N_2d_billiard_gamma} 
          (smooth solid line) shown over a larger energy range.               
          (c) Phase space at the lower horizontal boundary 
          with invariant region $\Gamma$ (red shaded), remaining phase space 
          (gray lines and dots), magnification of the hierarchical 
          regular-to-chaotic transition region, and illustration of a 
          regular trajectory.}
 \label{fig:weyl_reg_cosine}
\end{center}
\end{figure}

We now consider the desymmetrized cosine billiard 
in order to demonstrate that the partial Weyl law can be applied to systems with 
a generic mixed phase-space structure.
The cosine billiard is characterized by the height $h$ and length $a$ of the 
rectangular part as well as the height $h_c$ of the upper cosine boundary, see 
fig.~\ref{fig:weyl_reg_cosine}(c).
For the chosen parameters the phase-space of the cosine billiard consists of 
one large regular region surrounded by chaotic motion and a four-island 
resonance chain, see fig.~\ref{fig:weyl_reg_cosine}(c), which also shows a 
magnification of the generic hierarchical regular-to-chaotic transition region.
In order to apply the partial Weyl law we have to define a region $\Gamma$ 
in phase space and determine the corresponding area $A_\Gamma$ and length
$L_\Gamma$. According to eq.~\eqref{eq:N_2d_billiard_gamma} this gives a 
prediction for the number of eigenstates concentrating on $\Gamma$.
In principle any invariant region can be used. We choose the red-shaded region 
in fig.~\ref{fig:weyl_reg_cosine}(c), which contains most of the 
central regular island. The area $A_\Gamma$ is determined from
eq.~\eqref{eq:result_A_1} by numerical integration. 
For the length $L_\Gamma$ we
have to consider those parts of the billiard boundary $\partial\Omega$ for
which parallel trajectories belong to $\Gamma$. This holds for the
left vertical boundary of length $L_\Gamma=h+h_c$, 
see the inset in fig.~\ref{fig:weyl_reg_cosine}(a).
We stress, that including the hierarchical region or parts thereof in the 
definition of $\Gamma$ affects $A_\Gamma$, 
but not the boundary term $L_\Gamma$, as the orbits parallel 
to the boundary do not belong to the hierarchical transition region.
Numerically we calculate the first $1853$ eigenstates of the desymmetrized cosine
billiard with $h=0.8$, $a=1.3$, and $h_c=0.24$ using the improved method of 
particular solutions \cite{BetTre2005}. 
For the $l$th eigenstate we integrate its Poincar\'e-Husimi 
distribution \cite{BaeFueSch2004} over the region corresponding to
$\Gamma$ which gives the weight $w_{\Gamma}^{l}$. These weights determine 
the spectral staircase function $N_{\Gamma}(E)$.
We find excellent agreement with our prediction,
eq.~\eqref{eq:N_2d_billiard_gamma}, 
see fig.~\ref{fig:weyl_reg_cosine}(a). In
fig.~\ref{fig:weyl_reg_cosine}(b) we demonstrate that the boundary 
contribution of eq.~\eqref{eq:N_2d_billiard_gamma} is in agreement with the
difference of the numerical data and the first term of
eq.~\eqref{eq:N_2d_billiard_gamma}, apart from a constant 
offset due to higher order terms neglected in eq.~\eqref{eq:N_2d_billiard_gamma}.

\begin{figure}[b]
  \begin{center}
     \includegraphics[width=85mm]{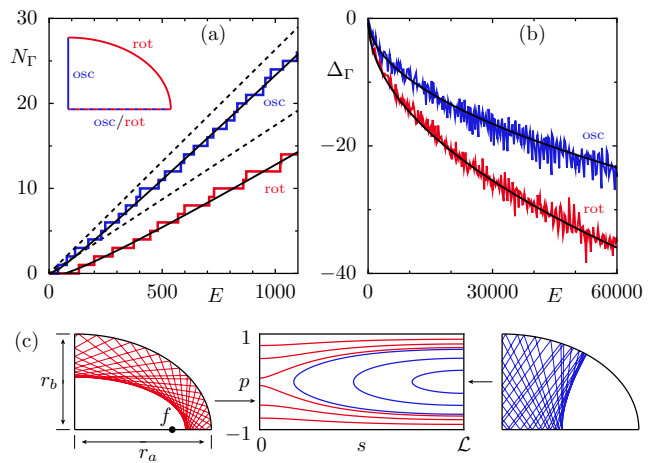}
 \caption{(Color online) (a) Spectral staircase $N_\Gamma(E)$
          for rotating and oscillating eigenstates of the 
          desymmetrized elliptical billiard with $r_a=1$ and $r_b=0.7$. 
          We compare numerical data
          with the first term (dashed lines) and both terms 
          of eq.~\eqref{eq:N_2d_billiard_gamma} (smooth solid lines).
          The inset shows the rotating and oscillating parts of the boundary.
          (b) Same data after subtracting the area term, 
          $\Delta_\Gamma(E) = N_\Gamma(E)-A_\Gamma E/(4\pi)$, 
          compared to the second term of eq.~\eqref{eq:N_2d_billiard_gamma} 
          (smooth solid lines) shown over a larger energy range.               
          (c) Phase space at the elliptical boundary 
          with rotating (red lines from left to right) and oscillating 
          (blue lines) regions and illustrations of trajectories.}
 \label{fig:weyl_reg_ellipse}
\end{center}
\end{figure}

As an interesting application, where a part of the boundary contributes to two 
invariant regions of phase space, we now consider the desymmetrized elliptical 
billiard \cite{WaaWieDull1997}, shown in fig.~\ref{fig:weyl_reg_ellipse}(c).
It is characterized by the lengths $r_a$ and $r_b$ of the two half-axes, 
with $r_a>r_b$, and the focus $f = \sqrt{r_{a}^{2} - r_{b}^{2}}$.
The phase space of the elliptical billiard consists of two separated regions 
of rotating and oscillating motion as visualized in 
fig.~\ref{fig:weyl_reg_ellipse}(c). 
The quantum eigenstates can be classified accordingly as mainly rotating or
oscillating, $\Nb(E) = \Nrot(E) + \Nosc(E)$. Using 
eq.~\eqref{eq:N_2d_billiard_gamma} we predict the number of rotating and 
oscillating states up to energy $E$. 
The areas $\Arot$ and $\Aosc$ are determined from
eq.~\eqref{eq:result_A_1} by numerical integration. For the length $\Lrot$ we
have to consider those parts of the billiard boundary $\partial\Omega$ for
which parallel trajectories show rotating motion. This is the case for the
elliptical boundary of length $\mathcal{L}$.  Trajectories parallel to the
horizontal boundary are precisely on the separatrix between oscillating and
rotating motion. Therefore, integration over $\delta(\sin\beta)$ in
eq.~\eqref{eq:res_d} gives half of the contribution for each of the two 
invariant regions of phase space and we have $\Lrot=\mathcal{L}+r_{a}/2$. 
For the oscillating states we have $\Losc=r_{b}+r_{a}/2$.
Note, that for the full ellipse the complete boundary belongs to the 
rotating region, $\Lrot=4\mathcal{L}$ and $\Losc=0$.
Numerically we calculate the first $2568$ eigenstates 
of the desymmetrized elliptical 
billiard with $r_a=1.0$ and $r_b=0.7$ using the improved method of 
particular solutions \cite{BetTre2005}. 
They are characterized by the angular and the radial quantum number 
$m$ and $n$. For each state we calculate the second constant of motion 
$\kappa_{mn}$ \cite{WaaWieDull1997}. If $\kappa_{mn}^2 > f^2$ the
state is classified as rotating and for $\kappa_{mn}^2 < f^2$ as oscillating.
Figure~\ref{fig:weyl_reg_ellipse}(a) shows the rotating and the oscillating 
spectral staircase for the elliptical billiard. 
We find excellent agreement with our prediction,
eq.~\eqref{eq:N_2d_billiard_gamma} (smooth solid lines). In
fig.~\ref{fig:weyl_reg_ellipse}(b) we demonstrate that the boundary 
contribution of eq.~\eqref{eq:N_2d_billiard_gamma} is in agreement with the
difference of the numerical data and the first term of
eq.~\eqref{eq:N_2d_billiard_gamma}. 
We have confirmed that also under variation of $r_b$ the prediction 
eq.~\eqref{eq:N_2d_billiard_gamma} agrees with numerics (not shown).

A straightforward generalization of our results to Neumann boundary 
conditions is possible by changing the sign of the second term in 
eqs.~\eqref{eq:dos_2d_billiard_gamma}, \eqref{eq:N_2d_billiard_gamma},
\eqref{eq:gf_2}, and \eqref{eq:res_d}.
As interesting tasks there remains to find the higher order terms of
$\db_\Gamma(E)$ due to corners and curvature effects as well as to generalize 
our approach to systems with broken time-reversal symmetry and to
three-dimensional cavities. Also the generalization of the approach 
to systems with smooth potentials \cite{BohTomUll1993} is an open problem. 
Finally, it is now possible to study the spectral fluctuations 
around $\db_\Gamma(E)$ associated with a phase-space region $\Gamma$  
for generic billiards.

We thank M.~Sieber for valuable discussions and
the DFG for support within the Forschergruppe 760 "Scattering Systems
with Complex Dynamics".


\begin{thebibliography}{30}
\bibitem[]{FOOTNOTE}
The standard Wigner function maps a plane wave to a sharp peak in
momentum. However, this clear correspondence is lost if the domain of the
wave function is restricted in space, as it is the case in a billiard. For a
point close to the boundary the integral in eq.~\eqref{eq:wigner} covers 
only a small spatial interval and results in a broad momentum distribution. 
The boundary adapted Wigner function avoids this artifact of the restricted domain 
\cite{BaeKetLoeSchFUTURE}. 
For example, in 1D the transform of a sine wave $\sin(\kappa q)$ has the expected
$\delta$-peaks at $p=\pm \kappa$. Note, that the standard Wigner function would result
in eq.~\eqref{eq:res_d} in a broad distribution of $\beta$ for points 
$\bq$ near the boundary.
\end{thebibliography}
\end{document}